\documentstyle[aps,prb,psfig]{revtex}
\newcommand{\beq}{\begin{equation}}
\newcommand{\eeq}{\end{equation}}
\newcommand{\eps}{\varepsilon} 
 
\begin{document} 
 
\author{J. O. Fj{\ae}restad and A. Sudb{\o}} 
 
\address{Department of Physics, Norwegian University of Science and  
Technology, N-7034 Trondheim, Norway.} 
 
\date{\today} 
 
\title{Non-k-diagonality in the interlayer pair-tunneling model \\ 
of high-temperature superconductivity} 
 
\maketitle 
                                                                               
\begin{abstract} 
We investigate the effect of $k$-space broadening of the interlayer
pairing kernel on the critical temperature $T_c$ and the
$k$-dependence of the gap function in a one-dimensional version of the
interlayer pair-tunneling model of high-$T_c$ superconductivity.  We
consider constant as well as $k$-dependent intralayer pairing kernels.
We find that the sensitivity to $k$-space broadening is larger the
smaller the width of the peak of the Fermi-level gap calculated for
zero broadening. This width increases with the overall magnitude of
the interlayer tunneling matrix element, and decreases with the
bandwidth of the single-electron intralayer excitation spectrum. The
width also increases as the Fermi level is moved towards regions where
the excitation spectrum flattens out. We argue that our qualitative
conclusions are valid also for a two-dimensional model.  This
indicates that at or close to half-filling in two dimensions, when the
Fermi-surface gap for zero broadening attains its peaks at $(\pm
\pi/a,0)$ and $(0,\pm\pi/a)$ where the excitation spectrum is flat,
these peaks should be fairly robust to moderate momentum broadening.
\end{abstract}

 


\section{INTRODUCTION} 
\label{intro} 
 
The interlayer pair-tunneling (ILT) model of high-temperature
superconductivity\cite{anderson} has been the focus of much attention
since it was introduced\cite{wheatley} and later elaborated on
quantitatively.  \cite{chakravarty1} Within the ILT model, the pairing
of electrons in individual CuO$_2$-layers is considerably enhanced by
the tunneling of Cooper pairs between neighbouring layers, giving
critical temperatures which are substantially higher than those
arising solely from a reasonable in-plane effective electron-electron
attraction in a two-dimensional (2D) BCS-like theory.

The central underlying assumption of the ILT model is that the normal
state of the cuprates is a strongly correlated non-Fermi liquid, where
{\em single}-electron interlayer tunneling is incoherent or strongly
damped, resulting in a frustrated $c$-axis kinetic energy.  This
kinetic energy is substantially lowered in the superconducting state,
through tunneling of Cooper pairs between CuO$_2$ layers.  Thus,
contrary to the situation in conventional superconductors, it is the
lowering of the kinetic energy, and not the potential energy, which
drives the transition.

Recently, there have been extensive discussions in the literature
about experimental tests of an unconventional relation, predicted by
the ILT model,\cite{anderson3,chakravarty3} between the $c$-axis
penetration depth $\lambda_c$ and the condensation energy
$E_{\rm{cond}}$.  The agreement in LSCO seems quite
good,\cite{uchida1,anderson3} but experiments on Hg-1201\cite{moler2}
and Tl-2201\cite{marel1,moler1,tsvetkov} give estimates for
$\lambda_c$ which are 8-20 times larger than the predicted
values.\cite{anderson3} However, note that Chakravarty {\em et
al.}\cite{chakkeab} have argued that this discrepancy between theory
and experiment can be drastically reduced by taking more carefully
into account the fluctuation contributions to the {\em normal state}
specific heat when estimating $E_{\rm{cond}}$.

It is not our purpose here to consider the microscopic foundations of
the ILT mechanism. Instead, we will take it as a phenomenological
starting point, and explore the effects of some modifications of the
form of the pair tunneling term used in
Ref. \onlinecite{chakravarty1}. There it was argued that, in order to
obtain critical temperatures of the same order of magnitude as found
in the high-$T_c$ cuprates, it was essential that the 2D momentum of
the Cooper-pair electrons was conserved in the tunneling process. This
momentum conservation was argued to follow from the momentum
conservation of the single-electron tunneling Hamiltonian, in the
absence of inelastic scattering.  Translated to real space, this
momentum conservation means that the electron-electron attraction
associated with the interlayer tunneling has an infinite range.
 
A natural question to ask is then how sensitive the critical
temperature $T_c$ is to a relaxation of this constraint. Specifically,
what is the typical order of the interaction range below which $T_c$
will drop to values which are no longer comparable to critical
temperatures in the high-$T_c$ cuprates? Moreover, several of the
unusual $k$-space features of the gap predicted within the ILT
mechanism have their origin in the assumed momentum conservation.

We will address this question phenomenologically, modelling the finite
range by postulating modified functional forms of the pair tunneling
term in which phenomenological parameters are introduced to measure
the degree of ``screening". This will in turn lead to modifications of
the original gap equations, which are then solved self-consistently to
obtain the critical temperature and the superconducting gap.  We
expect that {\it qualitatively} correct conclusions may be drawn from
our modelling of the k-space broadening. Brief accounts of parts of
this work have appeared in print elsewhere.\cite{yalta,sces98}

 
\section{FORMULATION OF THE PROBLEM} 
\label{formulation} 
 
For simplicity, we consider compounds with two CuO$_2$-layers per unit
cell.  The generalization to an arbitrary number of CuO$_2$-planes per
unit cell is straightforward.\cite{sudbo1} Below the superconducting
transition temperature, we will assume that the quasi-particle
description is approximately valid. The total Hamiltonian is taken to
be the sum of 2D BCS Hamiltonians for the individual layers, and an
interlayer pair tunneling Hamiltonian, $H=H_{\rm{layer}}+H_J$.  When
the zero-momentum pairing assumption is invoked, the intralayer part
is given by
\begin{eqnarray} 
H_{\rm{layer}}  =  \sum_{k,\sigma,i=1,2}\eps_k   
c^{(i)\dagger}_{k,\sigma} c^{(i)}_{k,\sigma}   
                +  \sum_{k,k',i=1,2} V_{k,k'}  
c^{(i)\dagger}_{k,\uparrow} c^{(i)\dagger}_{-k,\downarrow}  
c^{(i)}_{-k',\downarrow} c^{(i)}_{k',\uparrow}, 
\end{eqnarray} 
while the interlayer pair-tunneling contribution to the Hamiltonian is  
given by the form 
\beq 
H_J=-\sum_{k,k'}T_{J}(k,k')c^{(1)\dagger}_{k\uparrow}c^{(1)\dagger}_ 
{-k,\downarrow}c^{(2)}_{-k'\downarrow}c^{(2)}_{k'\uparrow}+\mbox{h.c.} 
\label{tjeq} 
\eeq 
Here $c^{(i)\dagger}_{k\sigma}$ is the creation operator of an
electron in layer $i$ ($i=1,2$) with 2D in-plane wave vector $k$ and
spin projection $\sigma$, $\eps_k$ is the normal state dispersion
measured relative to the Fermi level, and $V_{k,k'}$ is the inplane
contribution to the pairing kernel.
 
An apparently pathological aspect of a particular version of
(\ref{tjeq}), namely with a $k$-diagonal tunneling term $T_J(k,k')=
T_J \delta_{k,k'}$,\cite{chakravarty1} becomes evident on
Fourier-transforming back to real space, where it takes the form
\begin{eqnarray} 
-\frac{T_J}{N} \sum_{R_1,R_2,r}  
c^{(1)\dagger}_{R_1+r/2,\uparrow} c^{(1)\dagger}_{R_1-r/2,\downarrow} 
c^{(2)}_{R_2-r/2,\downarrow} c^{(2)}_{R_2+r/2,\uparrow} + \mbox{h.c.} 
\end{eqnarray}
where $N$ is the number of lattice sites per layer, and $r$ is the
relative coordinate and $R_i$ the center of mass coordinate in layer
$i$ of the two tunneling electrons . Note that there are no
restrictions on $|R_1-R_2|$ due to the zero-momentum pairing
assumption, as in conventional superconductors.  What is not
conventional is that there is no restriction on the relative positions
in each plane for which two electrons feel an attraction.\cite{foot1}
Hence, $T_J \delta_{k,k'}$ represents an infinite-range attraction,
contrary to the conventional case where it is a (retarded)
contact-attraction. That such a version of the ILT-model then gives a
large value of $T_c$ is perhaps not surprising, but it is difficult to
understand how such an effective attraction is produced.
 
The $k$-diagonal model must therefore be viewed as an idealization,
and the issue to adress is how representative this limit is, if at
all. The more general model given in (\ref{tjeq}) yields
\begin{eqnarray} 
-\frac{T_J}{N} \sum_{R_1,R_2,r} G(|r|) c^{(1)\dagger}_{R_1+r/2,\uparrow}  
c^{(1)\dagger}_{R_1-r/2,\downarrow} 
c^{(2)}_{R_2-r/2,\downarrow} c^{(2)}_{R_2+r/2,\uparrow} + \mbox{h.c.} 
\end{eqnarray} 
The characteristic decay-length of the function $G(|r|) =  
\sum_k e^{ikr}f(k)$, with $f(k)$ defined via $T_J(k,k')=T_J f(k-k')$,  
represents the range of the effective interlayer tunneling attraction.  
 
By assuming a layer-independent pair amplitude, the total Hamiltonian
becomes decoupled in the layer indices, and the gap equation is seen
to be the same as in the BCS case when one makes the replacement
$V_{k,k'}\to V_{k,k'}-T_J(k,k')$, i.e.  
\beq
\Delta_k=-\sum_{k'}V_{k,k'}\Delta_{k'} \chi_{k'}
+\sum_{k'}T_J(k,k')\Delta_{k'}\chi_{k'},
\label{origge} 
\eeq 
where $\Delta_k$ is the gap function, and $\chi_k$ is the pair
susceptibility, given by $\chi_k=\tanh(\beta E_k/2)/2E_k$, where
$E_k=\sqrt{\eps_k^2+|\Delta_k|^2}$, $\beta=1/k_B T$, $k_B$ is
Boltzmann's constant and $T$ is the temperature.  We will consider
$V_{k,k'}$ to be a separable function of $k$ and $k'$, i.e.
$V_{k,k'}= -V g_k g_{k'}$, where $g_{k}$ belongs to the set of basis
functions for irreducible representations of the point group of the
underlying lattice, and $V>0$ is an effective two-particle scattering
matrix element.


\section{Gap equation in energy space}
\label{gapen} 
 
Ref. \onlinecite{chakravarty1} studied the case
$T_J(k,k')=T_J\delta_{k,k'}$, i.e.  the pair tunneling matrix element
is both diagonal and $k$-independent.  Using also the BCS
approximation $g_k=\Theta(\omega_D-|\eps_k|)$, where $\omega_D$ is an
energy cutoff, the gap then depends on $k$ only through
$\varepsilon_k$, so that the gap equation can be written in energy
space as\cite{comment} 
\beq
\Delta(\eps)=\Delta_0\Theta(\omega_D-|\eps|)+T_J\Delta(\eps)\chi(\eps),
\label{EDIAG} 
\eeq 
where 
\beq
\Delta_0=\lambda\int_{-\omega_D}^{\omega_D}d\eps\,\Delta(\eps)\chi(\eps).
\label{delta0} 
\eeq 
The BCS coupling constant is $\lambda=V N(\eps_F)$, where
$N(\eps_F)$ is the density of states per spin at the Fermi level
$\eps_F$ (i.e. here we have made the usual approximation of neglecting
the variation of $\nabla_k \eps$ inside the thin Debye shell around
the Fermi energy).  This gap equation can be regarded as the limit
$\omega\to 0$ of the more general equation 
\beq
\Delta(\eps)=\Delta_0\Theta(\omega_D-|\eps|)+\frac{T_J}{2\omega}
\int_{\eps-\omega}^{\eps+\omega}d\eps'\,\Delta(\eps')\chi(\eps'),
\label{first} 
\eeq 
where the parameter $\omega$ provides a measure of the amount of $k$-space  
broadening in the interlayer pairing kernel. 

We have solved (\ref{first}) self-consistently and show in
Fig. \ref{figtcomega} the results for $T_c$ as function of $\omega$
for $T_J=30$ meV, $\omega_D=20$ meV and $\lambda=0.1$. The most
important feature of this figure is the moderate reduction of $T_c$ as
$\omega$ is increased from zero. To reduce $T_c$ by a factor $2$
requires a broadening of $\omega \sim 40$ meV.

If we convert the energy broadening of the ILT term to a length using
$\omega=\hbar^2 k^2/(2M)$, with $M$ equal to the electron mass, we
obtain for the length $l=1/k$ 
\beq l\approx
\left(\frac{62}{\sqrt{\omega}}\right)\mbox{\AA}, 
\eeq 
where $\omega$
is to be measured in meV. Setting $\omega=40$ gives an interaction
range $l\approx 9.8$ \AA.


\section{Gap equation in 1D $k$-space}

In this section, we will consider the gap equation (\ref{origge}) with
a particular choice of $T_J(k,k')$. The main purpose of this paper is
to establish a {\em qualitative} criterion for how robust the sharp
$k$-space structures of the gap, obtained for a $k$-diagonal ILT term,
are to momentum broadening. Given this limited purpose, it does make
sense to simplify the problem by taking the $k$'s to be
one-dimensional (1D). This simplification is purely mathematical, and
of course does not imply anything about superconductivity with true
off-diagonal long-range order in 1D systems, which is well-known not
to exist for $T>0$,\cite{hohenberg} and prohibited by quantum
fluctuations at $T=0$. The final justification of our 1D model lies in
the qualitative conclusions established at the end of this section,
which will be seen to apply also to a 2D system.

We will consider two different functional forms for $g_k$, both giving
a $k$-symmetric gap as required for singlet pairing. The first is the
BCS approximation $g_k=\Theta(\omega_D-|\eps_k|)$, also used in
Sec. \ref{gapen}. It is analogous to isotropic $s$-wave pairing in
2D. The second form is $g_k=\cos(ka)$, which is most closely analogous
to $s_{x^2+y^2}$ or $d_{x^2-y^2}$ pairing in 2D. The gap obtained for
the first form does not change sign in the Brillouin zone, while the
gap for the second form in general does.

For simplicity, we assume a simple tight-binding dispersion form for 
$\eps_k$, 
\beq 
\eps_k=-2t\left[\cos(ka)-\cos(k_F a)\right], 
\label{tbdisp} 
\eeq 
where $t$ is the single-electron intralayer tunneling matrix
element, $a$ is the lattice constant and $k_F$ is the Fermi wave
vector.  The pair tunneling term is taken to be of the form
$T_J(k,k')\equiv T_J f(k-k')$, where we have chosen $f(k)$ to have the
particular form 
\beq f(k)=\frac{k_0 a^2}{2L}\frac{1}
{\sin^2\left(\frac{ka}{2}\right)+\left(\frac{k_0a}{2}\right)^2},
\label{fkk} 
\eeq 
where $L$ is the length of the system and $k_0$ is a measure of
the width of $f(k)$. The prefactor in (\ref{fkk}) is chosen to ensure
a $k$-diagonal ILT term in (\ref{origge}) in the limit $k_0\to 0$. The
sine function ensures that the scattering is periodic in the
reciprocal lattice. One could construct infinitely many functions
$f(k)$ which reduce to a delta function as $k_0\to 0$, and hence our
particular choice (\ref{fkk}) is inevitably somewhat
arbitrary. However, since our focus here is merely on the qualitative
aspects of momentum broadening, the detailed form of $f(k)$ is of no
concern to us; {\em any} function $f(k)$ which is ``smeared out'' as
$k_0$ increases, would give the same qualitative results.  Note that
$G(r=0)=\sum_{k}f(k)=1/\sqrt{1+(k_0 a/2)^2}$, which means that the
effective value of $T_J$ actually decreases as $k_0 a$ is
increased. In this respect, the effect of momentum broadening is at
least not underestimated in our model.

\subsection*{Results and discussion} 
\label{results} 

We have calculated the critical temperature $T_c$ and the
zero-temperature gap function for various values of $k_0 a$ by solving
(\ref{origge}) self-consistently in the thermodynamic limit
$L\to\infty$. In Fig. \ref{figtck0} we show the results for $T_c$ for
$T_J=30$ meV, $\omega_D=20$ meV, $t=25$ meV, $k_F a=\pi/4$ and
$LV/2\pi a=2.5$ meV.  It is seen that $T_c$ is slightly more sensitive
to $k_0 a$ for $g_k=\cos(ka)$ than for
$g_k=\Theta(\omega_D-|\eps_k|)$. For $g_k=\cos(ka)$, $T_c$ is reduced
by a factor 2 compared to the $k$-diagonal result when $k_0
a/\pi\approx 0.25$. Only 1/10 of this broadening is required for a
50\% reduction of $T_c$ if one instead chooses $T_J=50$ meV, $t=250$
meV and $LV/2\pi a=25$ meV.\cite{sces98} The reason for this increased
sensitivity to broadening is the large increase of $t$.

In Fig. \ref{TJgap} we show the gap at $T=0$ for four values of $T_J$
and fixed $k_0=0$, the other parameter values being the same as used
for Fig. \ref{figtck0}.  In this case, the gap is given implicitly by
\beq 
\Delta_k=\frac{\Delta_0 g_k}{1-T_J\chi_k}, 
\eeq 
where $\Delta_0\equiv V\sum_k g_k\Delta_k\chi_k$.  The maximum of the 
gap, and hence the critical temperature $T_c$, is determined by $T_J$
through the enhancement factor $1/(1-T_J\chi_k)$, which has its
maximum on the Fermi surface. However, as seen in Fig. \ref{TJgap},
$T_J$ does not affect the sign of the gap, which is determined by
$g_k$ alone. On a 2D square lattice, the analogous statement is that
the transformation properties of the gap function under the symmetry
operations of the point group of the square lattice, $C_{4v}$, is
given entirely in terms of the intralayer contribution to the pairing
kernel,\cite{sudbo1,sudbo2} which is expandable in terms of basis
functions for the irreducible representations of $C_{4v}$.

In Fig. \ref{k0gap} we show the gap at $T=0$ for four values of $k_0
a$ and fixed $T_J=30$ meV. Note how the $k$-space variation of the gap
decreases with increasing $k_0 a$. For large enough $k_0 a$, $f(k)$ is
essentially independent of $k$, so the ILT term in (\ref{origge})
essentially becomes a constant self-consistent shift of $\Delta_k$.
The main contribution to the shift comes from the Fermi surface
region, where $\Delta_k$ and $\chi_k$ are maximal. Therefore, the sign
of the shift is essentially determined by the sign of $\Delta_k$ on
the Fermi surface, which in turn is determined by the sign of $g_k$ on
the Fermi surface, which for $g_k=\cos(ka)$ changes at
half-filling. Thus, the qualitative form of the gap is given by
$\Delta_k=\Delta_0 g_k+T_J\Delta_1$, where, for $g_k=\cos(ka)$, the
sign of $\Delta_1$ is positive below half-filling and negative above
half-filling. As a consequence of this shift, $\Delta_k$ eventually
ceases to change sign in the Brillouin zone for $g_k=\cos(ka)$, as
seen in Fig. \ref{k0gap}.

We now discuss the criterion for how much broadening is needed to
obtain a substantial reduction of the maximum value of the gap,
thereby smoothing out the sharp $k$-space structures obtained in the
$k$-diagonal case. For this purpose, it is instructive to consider how
a slightly broadened $T_J(k,k')$ affects the maximum value of the gap.
For $k_0 a/\pi \ll 1$, $\Delta_k$ varies more rapidly in the Fermi
surface region than $\chi_k$, because $1/(1-T_J\chi_k)$ is sharply
peaked at the Fermi surface. Thus the variation of $\Delta_k\chi_k$ in
the Fermi surface region is essentially determined by $\Delta_k$.
Furthermore, the main contributions to
$\sum_{k'}T_J(k_F,k')\Delta_{k'}\chi_{k'}$ roughly come from the
region $|k_F-k'|\lesssim k_0$. Temporarily denoting the gap calculated
for $k_0=0$ as $\Delta_k(0)$, it follows that as long as $k_0$ is much
smaller than the characteristic width of the peak of $\Delta_k(0)$,
the broadened $T_J(k_F,k')$ essentially has the same effect as a
$\delta$-function. Under such circumstances, the gap is little
affected by the non-$k$-diagonality. A broadening of the order of the
width of the peak of $\Delta_k(0)$ is therefore required for a
substantial effect of the broadening to be felt. Fig.  \ref{TJgap}
shows that the width of the peak of $\Delta_k(0)$ increases with
$T_J$. The detrimental effects on the gap of an increase of $k_0 a$
will therefore be reduced with an increase of $T_J$. On the other
hand, increasing $t$ will make the width of the peak of $\Delta_k(0)$
smaller, because the factor $1/(1-T_J\chi_k)$ drops more abruptly away
from its peak value as one moves away from the Fermi surface when the
overall amplitude of the variation of $\eps_k$ is increased, as seen
from the fact that this drop is proportional to 
\beq
\delta\varepsilon_k=2ta\sin(k_F a)\delta k.
\label{drop}
\eeq 
Thus the parameters $T_J$ and $t$ have opposite effects on the
sensitivity of the gap to broadening of $T_J(k,k')$. Note that one may
scale the parameter $t$ entirely out of (\ref{origge}) to obtain a gap
equation in terms of the dimensionless parameters $1/\beta t$,
$T_J/t$, $\Delta_k/t$, $V/t$ and $k_0 a$ (and $\omega_D/t$, when
$g_k=\Theta(\omega_D-|\eps_k|)$). It should also be mentioned that the
'realistic' values of $T_J/t$ are difficult to ascertain, because the
model we have considered is one-dimensional, and because the
experimentally relevant values of $T_J$ are hard to extract. For these
reasons, we can only draw qualitative conclusions from our model.

Another interesting consequence of (\ref{drop}) is that the width of
the peak of $\Delta_k(0)$ will increase as the Fermi level is moved
towards the band edges where the dispersion flattens out, since then
$\delta\varepsilon_k$ decreases. So in our 1D model the system becomes
more robust to a finite $k_0$ for a nearly empty or nearly full
conduction band.

We finally stress that our qualitative conclusions regarding the
sensitivity to momentum broadening are valid also for a 2D model. This
is because the arguments used to arrive at these conclusions depend on
premises that will be present also in 2D: 1) in the $k$-diagonal case,
the gap shows sharp enhancement at the Fermi surface, 2) the width
(and height) of the peak of this gap is increased by increasing the
amplitude $T_J$ of the interlayer tunneling matrix element, 3) there
will be parameters in the 2D single-electron intralayer dispersion
analogous to $t$ in the 1D case, which control the bandwidth of the
dispersion $\eps_k$, and therefore affect the width of the peak of the
$k$-diagonal gap in a manner similar to what occurs in the 1D
case. Note also that in the 2D case, the tight-binding dispersion
flattens out near the points $(\pm \pi/a,0)$ and $(0,\pm\pi/a)$, which
lie on the Fermi surface when the band is half-filled (for
nearest-neighbor hopping only) or close to half-filling (when
next-nearest neighbor hopping is included).\cite{thelen} These are
also the points where the $k$-diagonal gap is at its maximum for such
filling factors.\cite{chakravarty1} Thus it appears that near
half-filling in 2D, the maximum value of the gap should be fairly
robust to moderate momentum broadening.

 
\section{Conclusions} 

We have considered superconductivity within the ILT mechanism in the
presence of non-$k$-diagonal interlayer tunneling.  We find that the
sensitivity to momentum broadening is larger the smaller the width of
the peak of the gap obtained for $k$-diagonal tunneling. This width is
increased by increasing the amplitude $T_J$ of the interlayer
tunneling matrix element. The width is decreased by increasing the
bandwidth of the single-electron intralayer dispersion.  Finally, the
width is larger at points on the Fermi surface where the dispersion is
relatively flat as compared to points where the dispersion is
steeper.\cite{angilella} Although we illustrated these features by
solving a model with one-dimensional intralayer wavevectors, these
qualitative conclusions are also valid for the more experimentally
relevant case of two dimensions.
 
Several unusual properties of the superconducting state of the
cuprates are given an explanation with the ILT mechanism. The
essential feature of the ILT mechanism is the sharp $k$-space
structure of the gap that arises from an unusual enhancement factor
$1/(1-T_J \chi_k)$ for a k-diagonal interlayer tunneling. Conclusions
based on these sharp structures ought therefore to be reexamined in
the presence of a slightly broadened interlayer tunneling term.  This
pertains for instance to the explanation of the anomalies in the
neutron scattering peaks observed in YBCO using the ILT mechanism.  In
this case, non-trivial Fermi surface kinematics almost unique to the
mechanism are essential.\cite{chakravarty2}
 
\acknowledgments 
We thank N.-C. Yeh for useful discussions.
J.O.F. acknowledges support from the Norwegian University of  
Science and Technology through a university fellowship. Support from the  
Norwegian Research Council (Norges Forskningsr{\aa}d) through Grant No.  
110569/410 is also acknowledged. 





\begin{figure}
\caption{The critical temperature $T_c$ of the gap equation 
(\protect\ref{first}) as a function of the energy broadening $\omega$.}
\label{figtcomega}
\end{figure}
 
\begin{figure} 
\caption{
The critical temperature $T_c$ as a function of $k_0 a/\pi$.} 
\label{figtck0} 
\end{figure}

\begin{figure} 
\caption{ The $T=0$ gap function plotted for four values of $T_J$ with
$k_0=0$. The two cases $g_k=\Theta(\omega_D-|\eps_k|)$ and
$g_k=\cos(ka)$ are shown in the upper and lower panel,
respectively. As $T_J$ is increased, the maximum value of the gap,
occuring on the Fermi surface, increases and the variation of the gap
with $k$ is enhanced. Note how the sign of the gap is always
determined by $g_k$.}
\label{TJgap} 
\end{figure}

\begin{figure} 
\caption{ The $T=0$ gap function plotted for four values of $k_0
a/\pi$ with $T_J=30$ meV.  The two cases
$g_k=\Theta(\omega_D-|\eps_k|)$ and $g_k=\cos(ka)$ are shown in the
upper and lower panel, respectively. As $k_0$ is increased, the
maximum value of the gap and the variation of the gap with $k$
decreases. For large enough values of $k_0a/\pi$, this is reflected
for $g_k=\Theta(\omega_D-|\eps_k|)$ in a gap where the only $k$-space
variation comes from the sharp discontinuity in $g_k$, while for
$g_k=\cos(ka)$ the gap eventually ceases to change sign.}
\label{k0gap} 
\end{figure}

\begin{figure}
\psfig{file=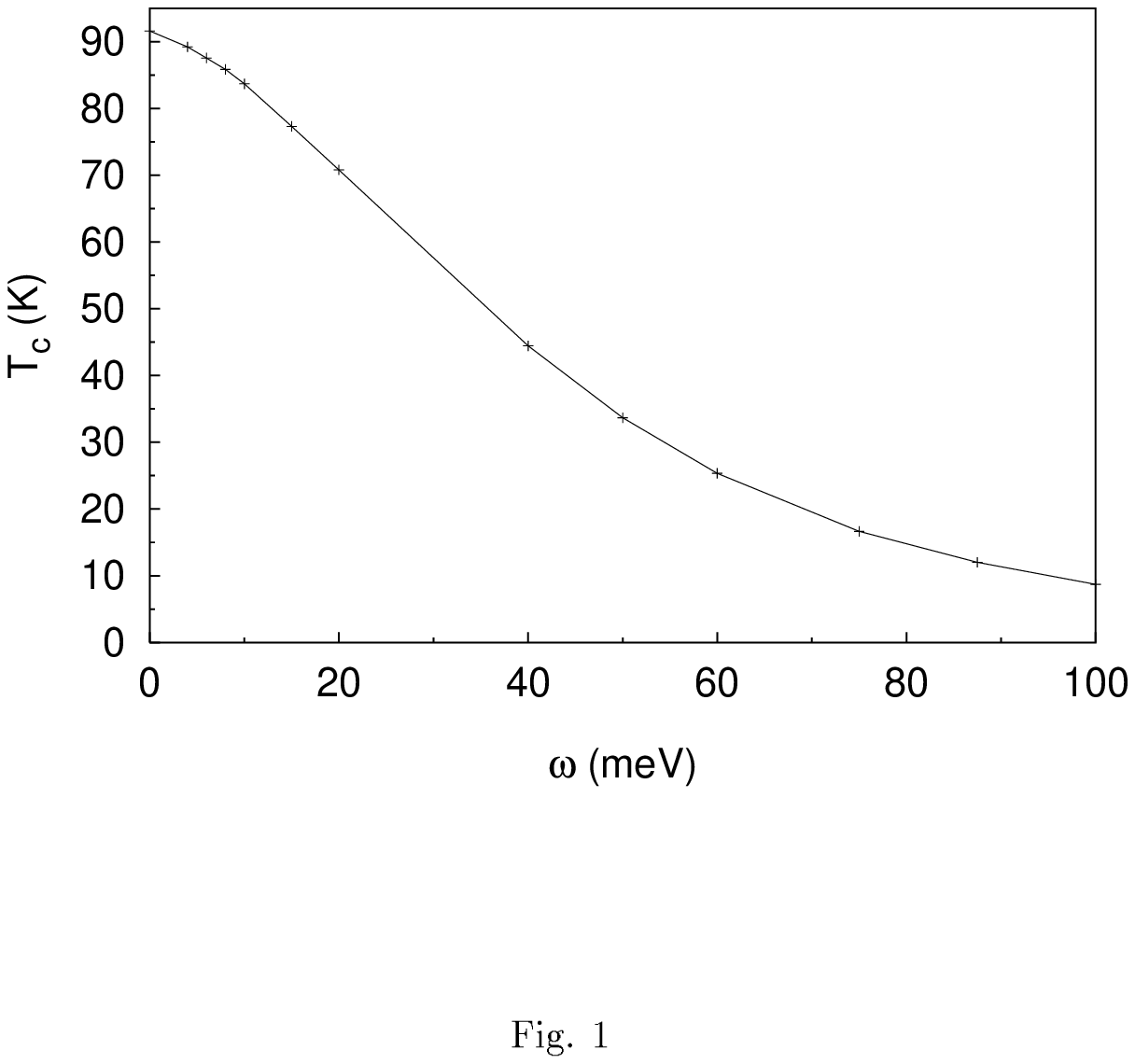,width=20cm}
\end{figure}

\begin{figure}
\psfig{file=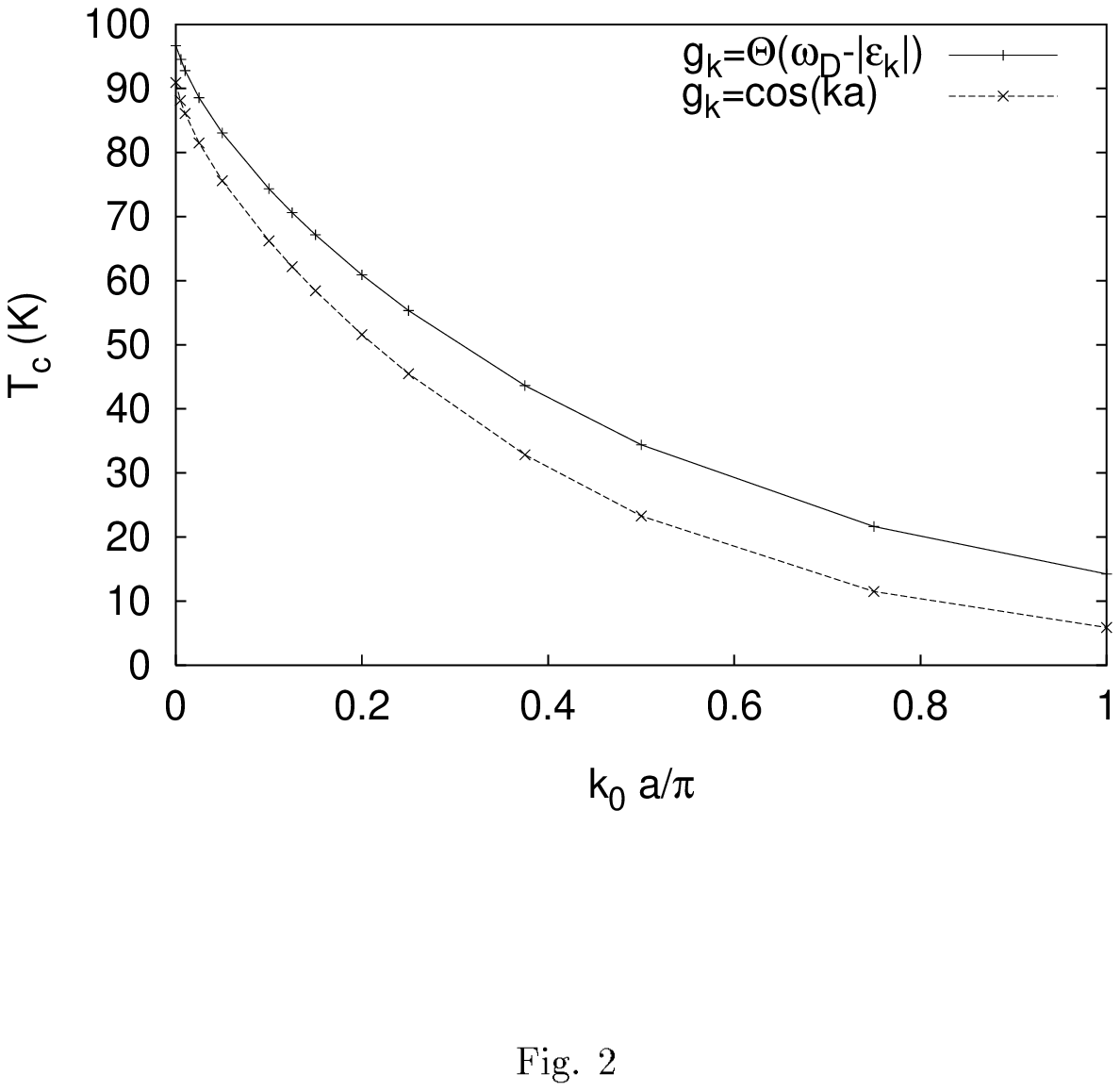,width=20cm}
\end{figure}

\begin{figure}
\psfig{file=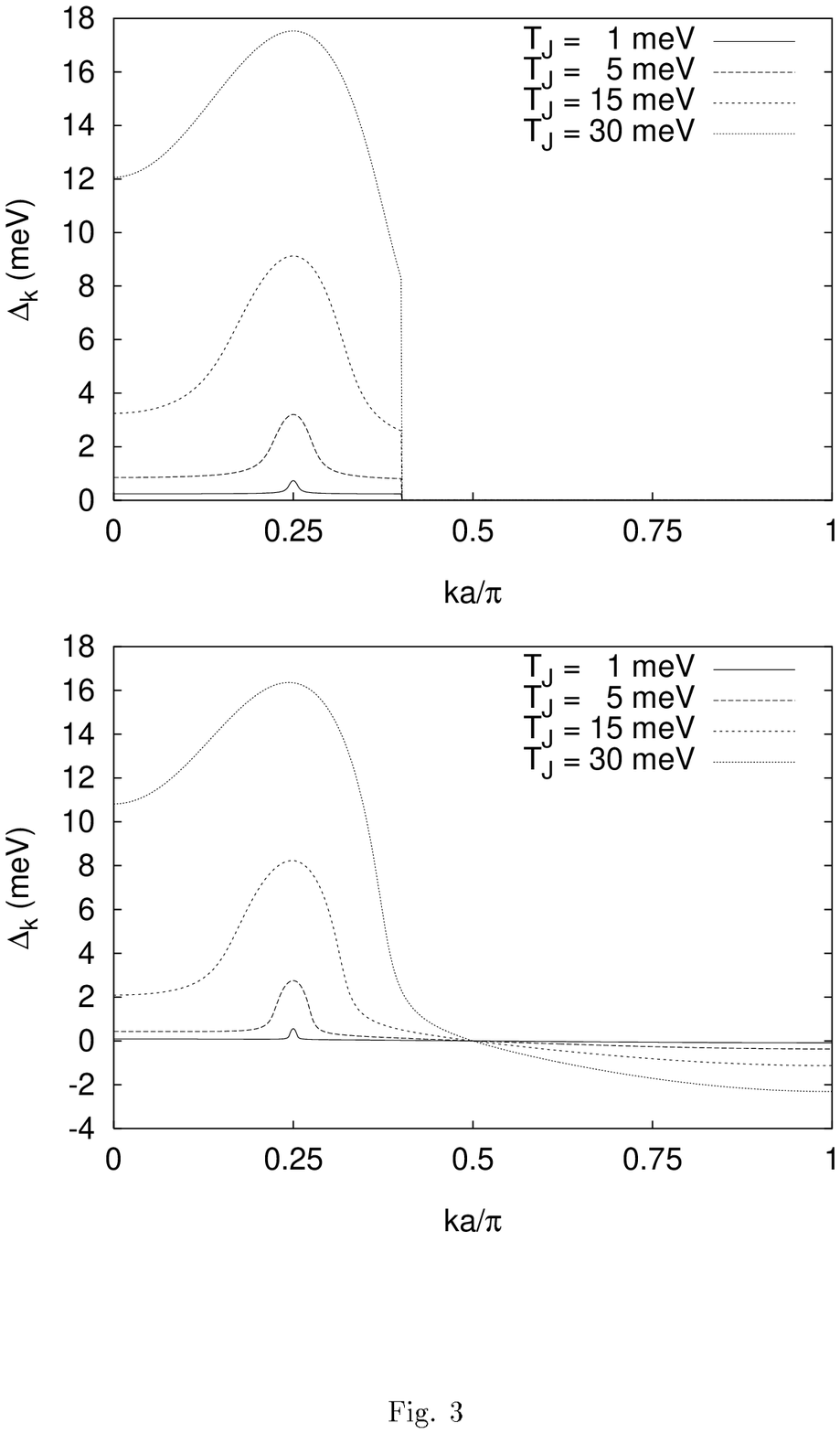,width=20cm}
\end{figure}

\begin{figure}
\psfig{file=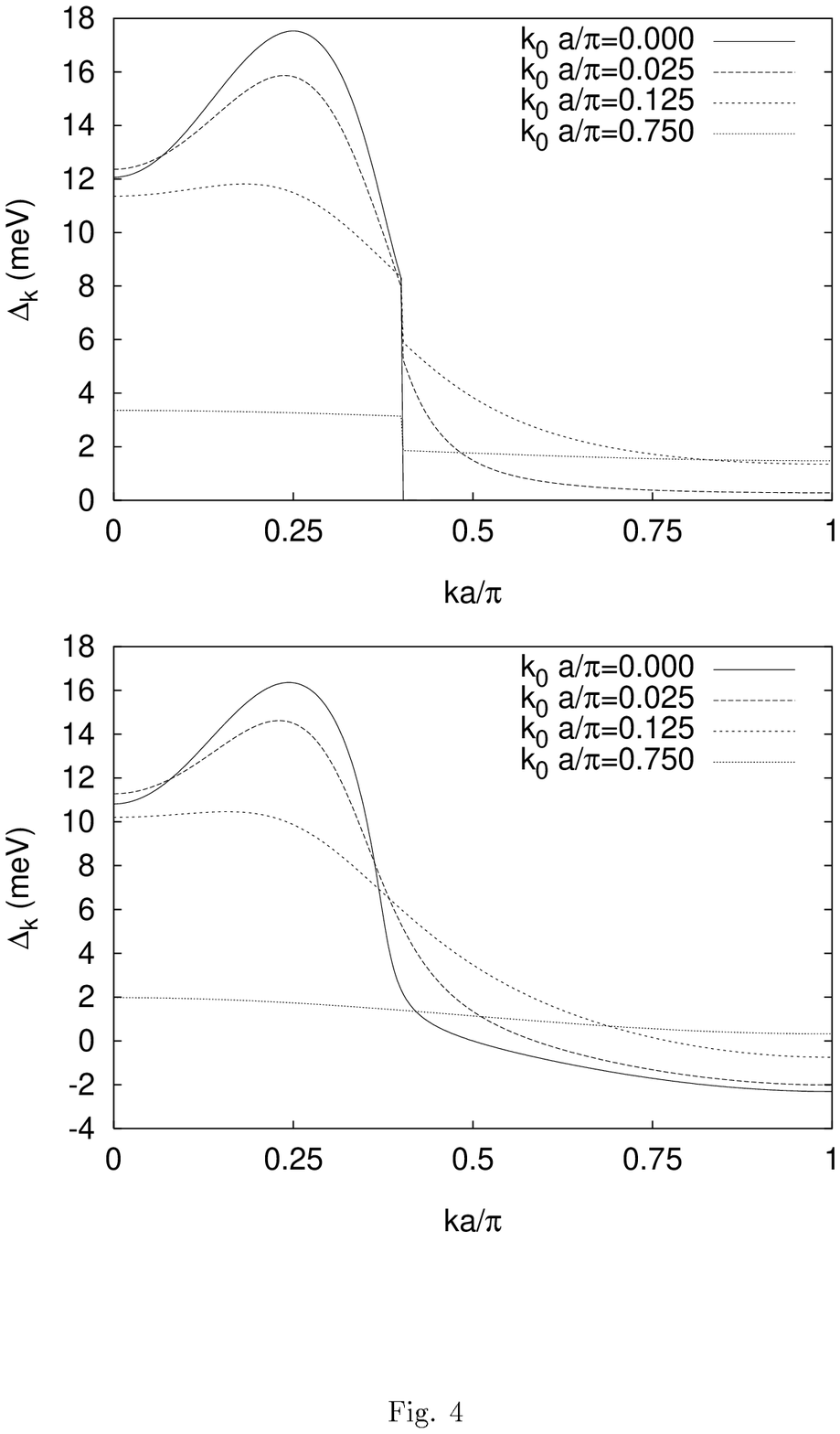,width=20cm}
\end{figure}


\begin{thebibliography}{99}
\bibitem{foot1} The relative positions of the tunneling electrons are
exactly conserved when $T_J(k-k') = T_J \delta_{k,k'}$, and $T_J$ is
taken to be a $k$-independent tunneling matrix element.  Using
instead\cite{chakravarty1} $T_J(k,k') = T_J(k) \delta_{k,k'}$ does not
alter the fact that there are no restrictions on the relative
positions for which two electrons may feel an effective
attraction. However, their relative positions are no longer exactly
conserved during the tunneling process.
\bibitem{comment} It was shown in Ref. \onlinecite{chakravarty1} that
(\ref{EDIAG}) gives a critical temperature $k_B T_c=T_J/4$ when the
intralayer contribution to the pairing kernel is ignored. Including
the contribution from the intralayer term gives, to lowest order, $k_B
T_c=(T_J/4)[1+3(\pi/2)^2\lambda^2]$, i.e. a quadratic $\lambda$
dependence.  In Ref. \onlinecite{chakravarty1} a linear dependence was
obtained, due to a specific approximation used for the pair
susceptibility which is excellent for $T_J=0$, but inaccurate for
$T_J\neq 0$.
\end{thebibliography}
\end{document}